\documentclass[aps, twocolumn, a4paper]{revtex4}

\usepackage{graphicx}

\begin{document}

\title{A new method to epitaxially grow long-range ordered\protect\\
       self-assembled InAs quantum dots on (110) GaAs}

\author{J.\ Bauer}
\author{D.\ Schuh}
\email{dieter.schuh@wsi.tu-muenchen.de}
\author{E.\ Uccelli}
\author{R.\ Schulz}
\author{A.\ Kress}
\author{F.\ Hofbauer}
\author{J.\ J.\ Finley}
\author{G.\ Abstreiter}
\affiliation{Walter Schottky Institut, Technische Universit\"at M\"unchen, Am Coulombwall 3, D-85748 Garching}

\begin{abstract}
We report on a new approach for positioning of self-assembled InAs
quantum dots on (110) GaAs with nanometer precision.
By combining self-assembly of quantum dots
with molecular beam epitaxy on in-situ cleaved surfaces
(cleaved-edge overgrowth) we have successfully fabricated
arrays of long-range ordered InAs quantum dots.
Both atomic force microscopy and micro-photoluminescence measurements
demonstrate the ability to control size, position, and ordering
of the quantum dots.
Furthermore, single dot photoluminescence investigations confirm
the high optical quality of the quantum dots fabricated.
\end{abstract}


\maketitle

The ability to precisely control the growth of self-assembled semiconductor quantum dots (QD)
is a topic that has recently attracted significant interest worldwide.
Today, In(Ga)As and In(Al)As dot layers in GaAs or AlAs can be grown
by molecular beam epitaxy (MBE) with
reasonably well defined emission energy, dot density and high size homogeneity
(e.g.\ \cite{ballet}).
However, to date it has been much
more difficult to
controllably position individual self-assembled quantum dots on the growth surface.
One possible approach is to exploit modified growth
kinetics that occurs on high index vicinal surfaces with regularly ordered atomic steps.
Examples of such novel approaches include In(Ga)As quantum dots and wires grown
on (311)A GaAs \cite{noetzel}, on miscut (100)-oriented
GaAs \cite{kim} or at crystal defects \cite{leon}.
Another approach is to use lithographically patterned substrates to force
controlled dot nucleation (e.g. \cite{petroff},\cite{schmidt}).
All of these methods are either based on intrinsic substrate properties and are,
therefore, not very flexible or are limited by the resolution of the lithographic technique used.
In the present work we demonstrate the ability to controllably
position quantum dots of a well defined size on a ($110$)-oriented surface
pre-structured with the atomic precision offered by MBE.
Atomic force microscopy and spatially resolved spectroscopy confirm the long-range
ordering and the excellent optical quality of these dots.
Our new approach to grow systems of aligned and well ordered quantum dots is to combine
self-assembly
with the cleaved-edge overgrowth \cite{pfeiffer}\ method.
In a first MBE step, a number of epitaxial layers were grown on a (001) GaAs
substrate. These precise structures act as a
template for quantum dot nucleation during a subsequent second MBE growth run on the cleaved
($1\bar{1}0$) surface.

\begin{figure}[h]
\includegraphics{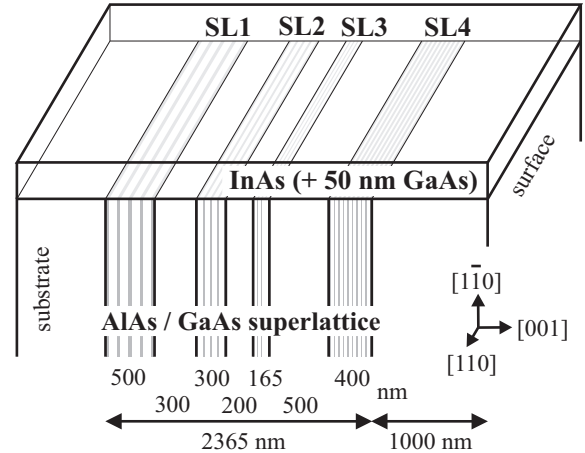}
\caption{%
\label{fig:fig1}
Schematic of the grown layer sequence:
In the first growth step the superlattices SL1--SL4
were grown on semi-insulating (001)-oriented GaAs.
The InAs dot layer and a 500~\AA\ thick capping layer
were grown in a subsequent MBE run on the ($1\bar{1}0$) GaAs
surface.}
\end{figure}

The sample investigated is depicted schematically in figure~\ref{fig:fig1} and consists of 4 spatially
separated AlAs/GaAs superlattices
(SL1: 5 periods of 320~\AA\ AlAs and 680~\AA\ GaAs;
 SL2: 5 periods of 200~\AA\ AlAs and 400~\AA\ GaAs;
 SL3: 5 periods of 110~\AA\ AlAs and 220~\AA\ GaAs;
 SL4: 10 periods of 200~\AA\ AlAs and 200~\AA\ GaAs)
grown on semi-insulating (001) GaAs.
The substrate temperature during growth was 650~$^\circ$C and the growth
rates of both GaAs and AlAs were 1~\AA/s under a III/V flux ratio
between 3 and 6.
Immediately after cleaving the substrate in the growth chamber,
we deposited 3 ML InAs on the ($1\bar{1}0$) surface.
For samples designed for atomic force microscopy (AFM) investigations,
growth was stopped after this step whereas
for photoluminescence measurements, the InAs layer was covered with 500~\AA\ GaAs
to bury the quantum dot
layer. Growth on the ($1\bar{1}0$) surface was performed at
470~$^\circ$C with an InAs growth rate of 0.1~\AA/s.

\begin{figure}[h]
\includegraphics{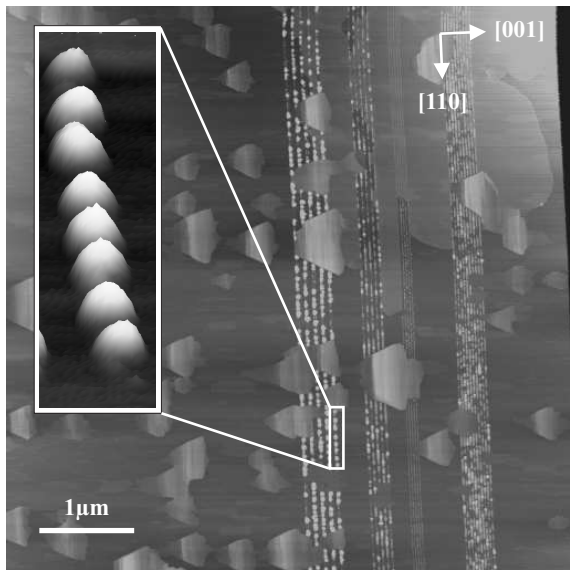}
\caption{%
\label{fig:fig2}
Atomic force microscopy picture of the well ordered
InAs quantum dots. The inset shows a close-up of 8 perfectly
aligned quantum dots with very similar size and shape. In the upper right corner of
the picture the edge of the
sample is visible.}
\end{figure}

Figure~\ref{fig:fig2} shows an AFM picture of the
uncapped sample.
Quantum dot-like nano\-struc\-tures can clearly be identified above all four superlattice regions
SL1--SL4 \cite{note:dotsnotseen}. Most surprisingly, the AFM measurements clearly indicate
nucleation of quantum dot-like nanostructures along the Al-rich regions of the exposed ($1\bar{1}0$)
growth surface. Furthermore, the typical size of the quantum dots is found to directly
reflect the thickness of the underlying AlAs layer;
being largest for SL1 (320~\AA\ AlAs), smallest for SL3 (110~\AA\ AlAs)
and comparable for SL2 and SL4 which share a common AlAs width (200~\AA) but differ in the thickness
of the surrounding GaAs and in the total number of periods.
However the quantum dots grown on SL2 and SL4 appear to have larger spatial
and structural inhomogeneities.
We conclude, therefore, that
the geometric properties of the quantum dots are sensitive not only to the thickness of the
underlying AlAs layer but also the surrounding GaAs
plays an important role in the dot nucleation process.
We suggest the following mechanism leading to
dot growth on the ($1\bar{1}0$) AlAs layers: due to the higher desorption rate of In atoms
on GaAs compared with AlAs and the lower In adatom mobility on Al-containing layers \cite{note:alasmobility}
there is an effective mass flow towards the AlAs layers. On top of these layers, the strain due to the
lattice mismatch between InAs and AlAs is reduced by the formation of a one-dimensional ordered
array of quantum dots.
The lower In adatom mobility on Al-rich surfaces may be the reason why
strain is not reduced by dislocations, as
known
for InAs growth on (110)-oriented GaAs \cite{belk}
but by nucleating in quantum dots.
In recent work, Wassermann {\em et al.\/} have also observed that the mechanism inhibiting
quantum dot formation on (110) surfaces could be overcome by InAs deposition on (110) AlAs
layers \cite{lyon}.

\begin{figure}[h]
\includegraphics{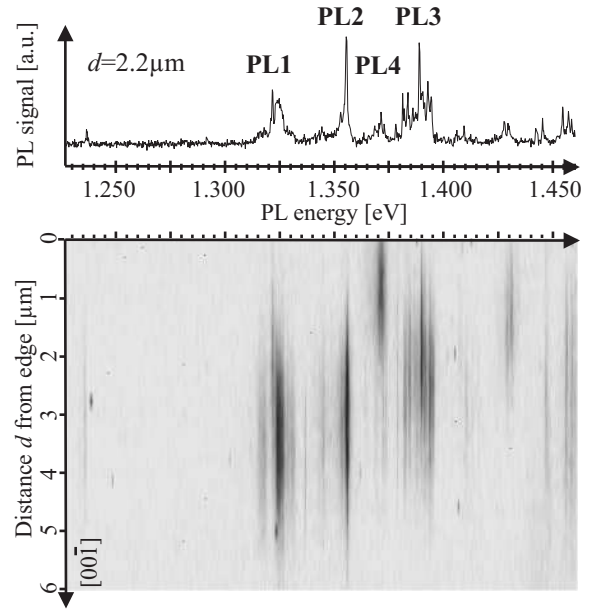}
\caption{%
\label{fig:fig3}
Grayscale map of $\mu$PL spectra
of the InAs quantum dots on the ($1\bar{1}0$) cleaved surface.
The spectra are taken along the [$00\bar{1}$] direction
at different distances $d$ from the edge of the sample.
The peaks between 1.3~eV and 1.4~eV are attributed to the buried
InAs quantum dot layer. In the upper panel, the spectrum recorded
at a distance $d$ = 2.2~$\mu$m is
presented for comparison.}
\end{figure}

In order to investigate the optical properties of the quantum dot nanostructures
we performed scanning $\mu$-photoluminescence ($\mu$PL) spectroscopy on a
subsequently grown sample with a GaAs capping layer.
The sample was mounted in a liquid He cryostat with the quantum dot layer accessible
for the excitation laser beam and the microscope to collect photoluminescence.
A HeNe laser, focused on the ($1\bar{1}0$) surface of the sample
was used to excite the quantum dots
and a confocal excitation and collection geometry provides a maximum spatial
resolution of
$\sim$1.5~$\mu$m.
The excitation spot was then raster scanned across the ($1\bar{1}0$) surface of the sample
to obtain a spatially and spectrally resolved image of the optical emission characteristics.
Spectra, taken at a temperature of $\sim$10 K and recorded at
different positions along the [$00\bar{1}$] direction are presented
in figure~\ref{fig:fig3} as a grayscale map.
In this figure, four different groups of sharp QD-like photoluminescence (PL)
peaks can be seen around 1.325~eV (PL1), 1.356~eV (PL2), 1.372~eV (PL4) and
1.389~eV (PL3).
By examining the spatial distribution of these luminescence peaks and correcting for
the spatial resolution of 1.5~$\mu$m we deduced the positions on the ($1\bar{1}0$) surface
of the underlying luminescence centers.
The positions obtained in this way
and the distance between the luminescence center near the edge of the sample
(PL4) and the most inner one (PL1) are in fairly good agreement
with the geometric properties of the superlattice substrate (see Fig.~\ref{fig:fig1}).
The PL signal labeled as group PL4 arises from a location next to the edge of the sample,
and the group PL3 appears next as we traverse the [$00\bar{1}$] direction towards bulk,
unstructured substrate.
Furthermore, the peaks of group PL3 show a higher PL energy
compared to those in group PL4.
Based on these observations we suggest that the PL signals of group PL4 belong to somewhat
larger QDs grown on SL4
while the blue-shifted PL signals of group PL3 belong to smaller QDs nucleated above SL3.
This suggestion is supported by the nominal
thickness of the AlAs layers in the respective superlattices
(SL3: 110~\AA, SL4: 200~\AA).
According to the energies of the PL signals in group PL1 and PL2, we suggest that
PL signals of group PL1 belong to the largest QDs grown on SL1
and that the PL signals of group PL2 belong to QDs nucleated above SL2.
Additionally, the locations of the maximum intensities of the relevant peaks
in these two groups are in very good agreement with the respective positions of SL1
and SL2 inside the sample.
Photoluminescence spectra recorded from positions further from the edge of the sample
revealed characteristically broader luminescence arising from the GaAs substrate but no
sharp-line emission in the spectral window below 1.45~eV due to quantum dots. This observation strongly
supports our identification of the sharp-line emission as arising from self-assembled
quantum dots which are spatially ordered into one-dimensional arrays along the AlAs
regions of the ($1\bar{1}0$) surface as shown in figure~\ref{fig:fig2}.

\begin{figure}[h]
\includegraphics{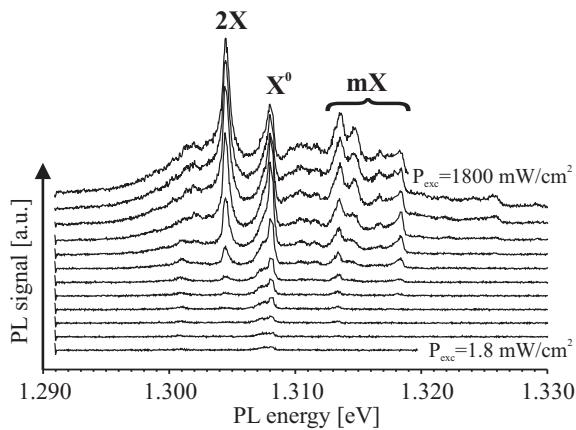}
\caption{%
\label{fig:fig4}
$\mu$PL spectra
of an single InAs quantum dot on the ($1\bar{1}0$) cleaved surface as a
function of excitation power.
For clarity, subsequent spectra are offset vertically.}
\end{figure}

In order to further substantiate our assumption that the sharp-emission lines are due to
laterally ordered self assembled quantum dots, we performed power dependent PL measurements
at a specific position of SL1.
The result of these measurements are summarized in figure~\ref{fig:fig4}.
At low excitation power, just a single emission line at 1.308~eV is
observed. The linear power dependence of this feature identifies it as
arising from a single exciton (X$^0$). In addition, this line is an emission doublet, possibly due to
elongation of the quantum dots along the AlAs layer and the resulting electron-hole
exchange interaction \cite{bayer}.
Upon increasing the excitation power density, several sharp lines at lower (1.3044~eV)
and higher (1.3135~eV, 1.3148~eV, 1.3166~eV and 1.3182~eV) energies emerge.
In particular, the intensity of the emission line at 1.3044~eV increases
quadratically with excitation power density and dominates the spectra for the highest
excitation densities investigated. This characteristic behavior identifies this peak
as arising from bi-exciton (2X) recombination in the dot \cite{brunner},
an observation which is further supported by the lineshape which,
in contrast with the single exciton, exhibits no exchange splitting as expected from the
spin-singlet nature of the bi-exciton ground state. The other lines arise from multi exciton
complexes (mX) and charged excitons \cite{findeis}.

In conclusion, we have demonstrated growth of
long-range ordered arrays of self-assembled quantum dots on ($110$) GaAs.
Both atomic force microscopy and photoluminescence experiments show that
precise control of size and order is possible due to the underlying AlAs/GaAs superlattices.
The fabricated quantum dots
show excellent optical properties and have a high potential for realisation of
well defined arrays of quantum dots with weak inhomogeneous broadening.
A detailed analysis of the optimised growth conditions is currently in progress.

This work was supported financially by Deutsche Forschungsgemeinschaft
via SFB 631 TP B1.


\begin{thebibliography}{99}
\bibitem{note:dotsnotseen}
 QDs on SL3 are to small to be resolved in this particular AFM picture shown here but
 were observed on the same sample by AFM with better resolution (not shown here).
\bibitem{note:alasmobility}
 this is known for InAs QD growth on (001)-oriented AlGaAs, see also Ref.\ \cite{ballet}
\end{thebibliography}
\end{document}